\documentclass[11pt]{article}
\usepackage{verbatim}
\usepackage{amsmath,amssymb}
\makeatletter \@addtoreset{equation}{section} \makeatother

\newcommand{\noi}{\vspace{12pt}\noindent}
\newcommand{\beq}{\begin{equation}}
\newcommand{\eeq}{\end{equation}}
\newcommand{\bea}{\begin{eqnarray}}
\newcommand{\eea}{\end{eqnarray}}

\newcommand{\e}[1]{{(\ref{#1})}}
\newcommand{\eq}[1]{{eq.\ (\ref{#1})}}
\newcommand{\es}[2]{{(\ref{#1}) and (\ref{#2})}}

\newcommand{\Ref}[1]{{Ref.~\cite{#1}}}

\newcommand{\equi}[1]{\stackrel{{#1}}{=}}

\newcommand{\ie}{{${ i.e., \ }$}}
\newcommand{\eg}{{${ e.g., \ }$}}

\newcommand{\cf}{{cf.\ }}

\newcommand{\wrt}{{with respect to }}

\renewcommand{\~}{ \ }
\renewcommand{\=}{ \ = \ }
\renewcommand{\Hat}{\widehat}
\renewcommand{\Tilde}{\widetilde}

\newcommand{\eps}{\varepsilon^{}}
\newcommand{\q}{{}^{}}
\newcommand{\p}{\!{}^{}}
\newcommand{\pp}{\!\!\!{}^{}}

\newcommand{\pl}{{\sf Pl}}
\renewcommand{\div}{{\rm div}}
\newcommand{\llbracket}{[\![}
\newcommand{\rrbracket}{]\!]}

\newcommand{\ad}{{\rm ad}}

\newcommand{\for}{{\rm for}}
\newcommand{\hf}{{\scriptstyle{\frac{1}{2}}}}
\newcommand{\Hf}{\frac{1}{2}}

\newcommand{\hi}{{\scriptstyle{\frac{\hbar}{i}}}}
\newcommand{\Ih}{\frac{i}{\hbar}}
\newcommand{\Hi}{\frac{\hbar}{i}}

\newcommand{\twostack}[2]{\begin{array}{c} \lower.8ex\hbox{${#1}$}
                     \cr \raise.8ex\hbox{${#2}$} \end{array}}
\newcommand{\deder}[1]{\frac{ 
 \stackrel{\raise.2ex\hbox{$\leftarrow$}}{\delta^{r}}   } 
 {   \delta {#1}}  }
\newcommand{\dedel}[1]{\frac{ 
 \stackrel{\lower.3ex \hbox{$\rightarrow$}}{\delta^{\ell}}   }
 {   \delta {#1}}  }
\newcommand{\papar}[1]{\frac{  
 \stackrel{\raise.2ex\hbox{$\leftarrow$}}{\partial^{r}}   } 
 {   \partial {#1}}  }
\newcommand{\papal}[1]{\frac{ 
 \stackrel{\lower.3ex \hbox{$\rightarrow$}}{\partial^{\ell}}   }
 {   \partial {#1}}  }
\newcommand{\rpa}[1]{{ 
 \stackrel{\raise.2ex\hbox{$\leftarrow$}}{\partial^{r}_{#1}}   }}
\newcommand{\lpa}[1]{{ 
 \stackrel{\lower.3ex\hbox{$\rightarrow$}}{\partial^{\ell}_{#1}}  }}
\newcommand{\larrow}[1]{\stackrel{\rightarrow}{#1}}
\newcommand{\rarrow}[1]{\stackrel{\leftarrow}{#1}}

\begin{document}
\thispagestyle{empty}
\title{\Large{\bf Gauge Independence in a Higher-Order Lagrangian Formalism \\
via Change of Variables in the Path Integral}}
\author{{\sc Igor~A.~Batalin}$^{a}$ and {\sc Klaus~Bering}$^{b}$ \\~\\
$^{a}$I.E.~Tamm Theory Division\\
P.N.~Lebedev Physics Institute\\Russian Academy of Sciences\\
53 Leninsky Prospect\\Moscow 119991\\Russia\\~\\
$^{b}$Institute for Theoretical Physics \& Astrophysics\\
Masaryk University\\Kotl\'a\v{r}sk\'a 2\\CZ--611 37 Brno\\Czech Republic}
\maketitle
\vfill
\begin{abstract}
In this paper we work out the explicit form of the change of variables that 
reproduces an arbitrary change of gauge in a higher-order Lagrangian formalism.
\end{abstract}
\vfill
\begin{quote}
PACS number(s): 03.70.+k; 11.10.-z; 11.10.Ef; 11.15.-q;  
\hrule width 5.cm \vskip 2.mm \noindent 
$^{a}${\small E--mail:~{\tt batalin@lpi.ru}} \hspace{10mm}
$^{b}${\small E--mail:~{\tt bering@physics.muni.cz}} \\
\end{quote}

\newpage

\section{Introduction}
\label{secintro}

\noi
It is a standard lore in the path integral formalism, that any result
(such as, \eg the Schwinger-Dyson equations, the Ward identities, etc.), that 
can be (formally) proven via change of integration variables, can 
equivalently be (formally) obtained via an integration by parts argument. 
And vice-versa. The latter method is typically the simplest. In 1996 it was 
shown in \Ref{bbd96}, by using integration by parts, how to formulate a 
higher-order field-antifield formalism that is independent of gauge choice. 
In this paper we work out the explicit form of the change of variables that 
reproduces a given change of gauge in an higher-order formalism.
Perhaps not surprisingly, the construction relies on identifying appropriate
homotopy operators.

\section{The $\Delta$ Operator}
\label{secdeltaop}

\noi
{}From a modern perspective \cite{witten90} the primary object in the 
Lagrangian field-antifield formalism \cite{bv81,bv83,bv84} is the 
{\bf $\Delta$ operator}, which is a nilpotent Grassmann-odd differential
operator 
\beq
\Delta^{2}\= 0\~,  \qquad \eps(\Delta)\=1\~,
\label{deltanilp01}
\eeq
and which depends on antisymplectic variables $z^{A}$ and their 
corresponding partial derivatives $\partial\p_{B}$. Their commutator\footnote{
The word {\em super} is often implicitly implied. {}For instance,
the word {\em commutator} means the supercommutator 
$[F,G]\equiv FG-(-1)^{\eps_{F}\eps_{G}}GF$.} reads
\beq
 \left[ \larrow{\partial}\pp_{B}, z^{A}\right] \= \delta^{A}_{B} \~.
\label{ccr01}
\eeq

\section{$Sp(2)$-Symmetric Formulation}
\label{secsp2deltaop}

\noi
We mention for completeness that there also exists an $Sp(2)$-symmetric 
Lagrangian field-antifield formulation \cite{bltla2}. This formulation
is endowed with two Grassmann-odd nilpotent, anticommuting $\Delta^{a}$ 
operators
\beq
\Delta^{\{a}\Delta^{b\}}\= 0\~,  \qquad \eps(\Delta^{a})\=1\~,  \qquad
a,b\~\in\~\{1,2\}\~.
\label{deltanilp02}
\eeq
Often (but not always!) the resulting $Sp(2)$-symmetric formulas look like 
the standard formulas with $Sp(2)$-indices added and symmetrized in a 
straighforward manner. In this paper, we will usually focus on the standard
formulation and only mention the corresponding $Sp(2)$-symmetric 
formulation when it deviates in a non-trivial manner.

\section{Planck Number Grading for Operators}
\label{secpl}

\noi
Planck's constant $\hbar$ is here treated as a formal parameter (as opposed 
to an actual number) in the spirit of deformation quantization (as opposed to
geometric quantization). 
The {\bf Planck number grading} $\pl$ for operators is defined via the rules
\beq
\pl(\hbar)\=1\~, \qquad \pl(z^{A})\=0\~, \qquad 
\pl(\larrow{\partial}\pp_{A})\=-1\~, \label{planchnumber01}
\eeq
and extended to normal-ordered\footnote{
{\em Normal-ordering} means that all the $z$'s appear to the left of all the 
$\partial$'s. {\em Antinormal-ordering} means the opposite. } 
differential operators in the natural way. More precisely, a derivative 
$\partial\p_{A}$ inside an operator $F$ gets assigned Planck number $-1$ 
($0$) for the parts that act outside (inside) the operator, respectively.
We mention for later convenience the superadditivity of Planck number grading
\beq
\pl(FG)\~\geq\~\pl(F)+\pl(G)\~, \qquad 
\pl([F,G])\~\geq\~\pl(F)+\pl(G)+1\~,
\label{planchnumbercom01}
\eeq
where the uppercase letters $F$ and $G$ denote operators.

\section{Higher-Order $\Delta$ Operator}
\label{sechighdeltaop}

\noi
In the standard field-antifield formalism \cite{bv81,bv83,bv84}, 
the $\Delta$ operator is a second-order operator. (See also 
Section~\ref{sec2ndorder}.) 
In the higher-order generalization \cite{bbd96}, which is the main topic of 
this paper, the $\Delta$ operator is assumed to have Planck number grading
\cite{bt94def} 
\beq
\pl(\Delta)\~\geq\~ -2\~.\label{triangularform01}
\eeq
Evidently, the Planck number inequality \e{triangularform01} means that the 
normal-ordered $\Delta$ operator is of the following triangular form\footnote{
In contrast to the original proposal \cite{bbd96}, we also allow the three 
terms $\Delta\q_{-2,0}$, $\Delta\q_{-1,0}$ and $\Delta\q_{-1,1}$ with negative 
$n$ in \eq{triangularform02}. The two last terms arise naturally in the
$Sp(2)$-symmetric formulation \cite{bltla2,bm96}. The two first terms 
affect the classical master eq. See also 
Sections~\ref{sec2ndorder}-\ref{sec2ndorderappl} for the second-order case.}
\beq
\Delta \=\sum_{n=-2}^{\infty}\sum_{m=0}^{n+2}\left(\hi\right)^{n}\Delta\q_{n,m}\~, 
\qquad\Delta\q_{n,m}\=
\Delta_{n,m}^{A\q_{1}\ldots A\q_{m}}(z)\larrow{\partial}\pp_{A\q_{m}}\ldots 
\larrow{\partial}\pp_{A\q_{1}}  \~.\label{triangularform02}
\eeq
The higher-order terms in the $\Delta$ operator can, \eg be physically
motivated as quantum corrections, which arise in the correspondence 
between the path integral and the operator formalism.

\section{Path Integral}
\label{secpathint}

\noi
The (formal) path integral 
\beq
Z\p_{X}\= \int \! d\mu \~wx\~, \qquad w\~\equiv\~e^{\Ih W}\~, 
\qquad x\~\equiv\~e^{\Ih X}\~,  \label{pathint01} 
\eeq
in the $W$-$X$-formalism \cite{bt92,bt93,bt94,bms95,bm96,bt96,bbd06} 
consists of three parts:
\begin{enumerate}
\item
A {\bf path integral measure} $d\mu=\rho [dz][d\lambda]$, where 
$\lambda^{\alpha}$ are Lagrange multipliers implementing the gauge fixing 
conditions, and $z^{A}\equiv\{\phi^{\alpha}; \phi^{\ast}_{\alpha}\}$ are the 
antisymplectic variables, \ie fields $\phi^{\alpha}$ and antifields
$\phi^{\ast}_{\alpha}$. Here $\rho=\rho(z)$ is a density with 
$\eps(\rho)=0$ and $\pl(\ln\rho)\geq -1$.
\item
A gauge-generating {\bf quantum master action} $W$, which satisfies 
the {\bf quantum master equation} (QME)\footnote{
The parenthesis in \eq{qmew} is here meant to emphasize that the QME is 
an identity of functions (as opposed to differential operators), \ie the 
derivatives in $\Delta$ do {\em not} act outside the parenthesis. Note however 
that similar parenthesis will not always be written explicitly in order not 
to clog formulas. In other words, it must in general be inferred from the 
context whether an equality means an identity of functions or an identity of 
differential operators.}
\beq
(\Delta w)\= 0\~, \qquad w\~\equiv\~e^{\Ih W}\~, \qquad \pl(W)\~\geq\~0\~.
\label{qmew}
\eeq 
The path integral \e{pathint01} will in general depend on $W$, since $W$
contains all the physical information about the theory, such as, \eg the
original action, the gauge generators, etc.\ \cite{bv85,bb10}.
The triangular form \e{triangularform02} of the $\Delta$ operator implies
that the QME \e{qmew} is perturbative in Planck's constant $\hbar$, \ie
\beq
\pl \left( w^{-1}\Delta(\hbar,z,\partial) w \right)\=
\pl \left( \Delta \left(\hbar,z,\partial+\Ih (\partial W)\right)  \right) 
\~\geq\~-2\~.\label{planckexplained01}
\eeq
Besides the triangular form \e{triangularform01}, which is imposed to ensure 
perturbativity, there are additional ``boundary'' and rank conditions to 
guarantee the pertinent classical\footnote{
The word {\em classical} means here independent of Planck's constant $\hbar$.} 
master equation and proper classical master action $S$ \cite{bv85,bb10}.

\item
A gauge-fixing {\bf quantum master action} $X$, which satisfies the transposed
quantum master equation
\beq
(\Delta^{T}x)\= 0\~, \qquad x\~\equiv\~e^{\Ih X}\~, \qquad \pl(X)\~\geq\~0\~.
\label{qmex}
\eeq
The path integral \e{pathint01} will in general not depend on $X$, \cf 
Section~\ref{secgaugeindepintbyparts} and Section~\ref{secgaugeindepchange}.
\end{enumerate}

\noi
The {\bf transposed operator} $F^{T}$ has the property that
\beq
\int \! d\mu\~ (F^{T} f)\~ g
\= (-1)^{\eps_{f}\eps_{F}}\int \! d\mu\~ f\~ (F g)\~. 
\label{transposed01}
\eeq
Here the lowercase letters $f,g, \ldots$ denote functions, while the upper
case letters  $F,G, \ldots$ denote operators. One can construct any 
transposed operator by successively apply the following
rules 
\beq(F+G)^{T}\=F^{T}+G^{T}\~,\qquad
(FG)^{T}\=(-1)^{\eps_{F}\eps_{G}}G^{T}F^{T}\~,\qquad  
(z^{A})^{T}\=z^{A}\~, \qquad \partial^{T}_{A} \=-\rho^{-1}\partial\p_{A}\rho\~.
\label{transposed02}
\eeq
In particular the transposed operator $\Delta^{T}$ is also nilpotent
\beq
\left(\Delta^{T}\right)^{2}\= 0\~.
\label{deltatnilp01}
\eeq
The transposed derivative $\partial^{T}_{A}$ satisfies a modified Leibniz rule:
\beq
 \partial^{T}_{A}(fg)\=  (\partial^{T}_{A}f) g 
- (-1)^{\eps_{A}\eps_{f}}f (\partial\p_{A}g)\~. \label{affineleibniz}
\eeq
Let us mention for completeness that the $\Delta$ operator (which takes 
functions to functions) and the $W$-$X$-formalism can be recast in terms of 
Khudaverdian's operator $\Delta\q_{E}$ (which takes semidensities to 
semidensities) \cite{k99,kv02,k02,k04,b06,b07,bb07,bb08,bb09}.

\section{Higher-Order Quantum BRST Operators}
\label{secbrst}

\noi
The quantum BRST operators $\sigma\q_{W}$ and $\sigma\q_{X}$ take operators 
into functions (\ie left multiplication operators). They are defined as
\beq
 \sigma\q_{W} F \~:=\~ \Hi w^{-1} ([\Delta,F] w)
\~\equi{\e{qmew}}\~ \Hi w^{-1} (\Delta Fw)\~, \label{sigmaw01}
\eeq
\beq
 \sigma\q_{X} F \~:=\~  \Hi x^{-1} ([\Delta^{T},F] x)
\~\equi{\e{qmex}}\~ \Hi x^{-1} (\Delta^{T}Fx)\~,\label{sigmax01}
\eeq
respectively, where $F$ is an operator. They are nilpotent, Grassmann-odd,
\beq
\sigma^{2}_{W}\=0\=\sigma^{2}_{X}\~, \qquad  
\eps(\sigma\q_{W})\=1\=\eps(\sigma\q_{X}) \~,\label{sigmanilp01}
\eeq
and perturbative in the sense that
\beq
\pl\left(  \sigma\q_{W} F \right) \~\geq\~ \pl(F)
\~\leq\~ \pl\left(  \sigma\q_{X} F \right)\~.\label{sigmaperp01}
\eeq
In the $Sp(2)$-symmetric formulation the quantum BRST operators 
$\sigma^{a}_{W}$ and $\sigma^{a}_{X}$ carry an $Sp(2)$-index since the
$\Delta^{a}$ operator does.

\section{Higher quantum antibrackets}
\label{sechighqantibrackets}

\noi
The {\bf $1$-quantum antibracket} is defined as
\beq
\Hat{\Phi}_{\Delta}^{1}\~\equiv\~D\~\equiv\~ [\Delta,\~\cdot\~]\~, \qquad  
D^{2}\~\equiv\~0\~, \qquad \eps(D)\= 1\~, \qquad \pl(DF)\~\geq \~\pl(F)-1\~.
\label{dee01}
\eeq
The {\bf $n$-quantum antibracket} consists of nested commutators of $n$ 
operators with the $\Delta$-operator 
\cite{yks96,yks04,bm98,bm99,bm99dual,bm99nonilp,b06cmp}. We will not need 
the full definition here, but it can in principle be deduced uniquely via 
polarization of the following recursive formula \cite{b06cmp} 
\beq
\Hat{\Phi}_{\Delta}^{n}\!\underbrace{(B,\ldots, B)}_{n \text{ arguments}}
\~:=\~ 
[\ldots [[\Delta,\underbrace{B],B], \ldots,B]}_{n \text{ arguments}}
\=[\Hat{\Phi}_{\Delta}^{n-1}\!\underbrace{(B,\ldots, B)}_{n-1 \text{ arguments}},B]\~,
\qquad \eps(B)\=0, \qquad \Hat{\Phi}_{\Delta}^{0}\=\Delta  \~.\label{phin01}
\eeq
Philosophically speaking, the $n$-quantum antibrackets \e{phin01} are 
secondary/derived objects, which can be obtained from the 
underlying concept of a fundamental $\Delta$-operator. 
The pertinent Lie bracket is the {\bf $2$-quantum antibracket} / derived 
bracket
\cite{yks96,yks04,bm98}
\beq
\llbracket F, G\rrbracket
\~:=\~\Hf [F, D G] - (-1)^{(\eps_{F}+1)(\eps_{G}+1)}(F \leftrightarrow G)
\=-(-1)^{\eps_{F}}\Hat{\Phi}_{\Delta}^{2}(F,G)\~,\label{qantibracket01}
\eeq
where
\beq
\Hat{\Phi}_{\Delta}^{2}(F,G)
\~:=\~\Hf [D F,G]+ (-1)^{\eps_{F}\eps_{G}} (F \leftrightarrow G)\~.
\label{phi201}
\eeq
The $2$-quantum antibracket is Grassmann-odd
\beq
\eps\left(\llbracket F, G\rrbracket\right) \= \eps(F)+\eps(G)+1\~, 
\label{phi2odd01}
\eeq
and perturbative 
\beq
 \pl\left(\llbracket F, G\rrbracket \right)
\~\geq\~ \pl(F)+\pl(G)\~.\label{phi2perp01}
\eeq
The $1$-quantum antibracket  $D$ generates the $2$-quantum antibracket
\cite{bm98}
\beq
 [D F, D G]
\=D \llbracket F, G\rrbracket 
\= \llbracket D F, G\rrbracket
-(-1)^{\eps_{F}} \llbracket F ,D G\rrbracket \~. \label{deedee01}
\eeq
The $3$-quantum antibracket is defined as
\beq
\Hat{\Phi}_{\Delta}^{3}(\Psi\q_{1},\Psi\q_{2},\Psi\q_{3})
\~:=\~\frac{1}{6} \sum_{{\rm cycl}.\ 1,2,3}
[[D\Psi\q_{1},\Psi\q_{2}],\Psi\q_{3}] 
- (1\leftrightarrow 2)\~,\qquad \eps(\Psi\q_{i})\=1\~.\label{phi301}
\eeq
The Jacobi identity for the $2$-quantum antibracket is satisfied up to 
$D$-exact terms
\beq
\sum_{{\rm cycl}.\ 1,2,3}\llbracket\llbracket\Psi\q_{1},
\Psi\q_{2}\rrbracket, \Psi\q_{3}\rrbracket
\= \Hf D\Hat{\Phi}_{\Delta}^{3}(\Psi\q_{1},\Psi\q_{2},\Psi\q_{3})
\~,\qquad \eps(\Psi\q_{i})\=1  \~, \label{jacid01a}
\eeq
or equivalently, in the polarized language \cite{b06cmp} 
\beq
6 \Hat{\Phi}_{\Delta}^{2}(\Hat{\Phi}_{\Delta}^{2}(B,B),B)
\=  D\Hat{\Phi}_{\Delta}^{3}(B,B,B)
\~,\qquad \eps(B)\=0  \~. \label{jacid01b}
\eeq
{\sc Proof of \eq{jacid01b}}: 
\bea
&&6 \Hat{\Phi}_{\Delta}^{2}(\Hat{\Phi}_{\Delta}^{2}(B,B),B)
-D\Hat{\Phi}_{\Delta}^{3}(B,B,B)
\= 3[DB,[DB,B]] + 3 [D[DB,B],B] - D[[DB,B],B] \cr
&=& 4[[DB,B],DB] + 2 [D[DB,B],B]  
\= 4[DB,[DB,B]] - 2 [[DB,DB],B]\=0 \~. \label{jacid01c}
\eea

\section{Grassmann-even $Sp(2)$ quantum brackets}
\label{sechighqbrackets}

\noi
In the $Sp(2)$-symmetric case, besides the $Sp(2)$-symmetric higher quantum
antibrackets (which we will not discuss here), there is a tower of 
Grassmann-even quantum brackets.
The pertinent {\bf $1$-quantum bracket} is
\beq
D\~\equiv\~
\Hf\Hi\epsilon\q_{ab}[\Delta^{a},[\Delta^{b},\~\cdot\~]]
\~, \qquad  D^{2}\~\equiv\~0\~, \qquad \eps(D)\= 0\~, \qquad 
\pl(DF)\~\geq \~\pl(F)-1\~.\label{dee02}
\eeq 
The {\bf $2$-quantum bracket} is defined as
\beq
\llbracket F, G \rrbracket
\~:=\~ \Hf [DF, G] + \Hf [ F, DG] 
\= -(-1)^{\eps_{F}\eps_{G}}\llbracket G, F \rrbracket \~.
\label{qbracket02}
\eeq
(Hopefully it does not lead to confusion that we use the same notation for the
Grassmann-even quantum brackets $D$ and $\llbracket \cdot, \cdot \rrbracket$ 
in this Section~\ref{sechighqbrackets} as we use for the Grassmann-odd quantum 
antibrackets $D$ and $\llbracket \cdot, \cdot \rrbracket$ in the previous 
Section~\ref{sechighqantibrackets}.) Up to $D$-exact terms, the $2$-quantum 
bracket is
\beq
\llbracket F, G\rrbracket-\Hf D[F, G]
\= \frac{1}{4}\Hi \epsilon\q_{ab}
[[ F, \Delta^{a}], [\Delta^{b} ,G]] 
-(-1)^{\eps_{F}\eps_{G}} (F \leftrightarrow G)\~,
\label{phi202}
\eeq
The $2$-quantum bracket is Grassmann-even
\beq
\eps\left(\llbracket F, G\rrbracket\right) \= \eps(F)+\eps(G)\~, 
\label{phi2even02}
\eeq
and perturbative 
\beq
 \pl\left(\llbracket F, G\rrbracket \right)
\~\geq\~ \pl(F)+\pl(G)\~.\label{phi2perp02}
\eeq
The $1$-quantum bracket $D$ generates the $2$-quantum bracket 
\beq
 [D F, D G]
\=D \llbracket F, G\rrbracket 
\= \llbracket D F, G\rrbracket
+ \llbracket F ,D G\rrbracket \~. \label{deedee02}
\eeq
We note for later the identity
\beq
[F,DG]-[DF,G]\=
D[F,G] - \Hi \epsilon\q_{ab} [\Delta^{a}, [[\Delta^{b},F],G]]\~. 
\label{weird01}
\eeq
The Jacobi identity for the $2$-quantum antibracket is satisfied up to 
$D$-closed terms
\beq
\sum_{{\rm cycl}.\ 1,2,3}\llbracket\llbracket B\q_{1},
B\q_{2}\rrbracket, B\q_{3}\rrbracket
\~\sim\~ 0  \~,\qquad \eps(B\q_{i})\=0\~.
\label{jacid02a}
\eeq
In detail, in the polarized language \cite{b06cmp} 
\beq
6 \llbracket\llbracket\Psi,\Psi\rrbracket, \Psi\rrbracket
\= D[\llbracket\Psi,\Psi\rrbracket, \Psi] 
+ \Hi \epsilon\q_{ab} [\Delta^{a},[[\Delta^{b},\Psi],
\llbracket\Psi,\Psi\rrbracket]]
\~,\qquad \eps(\Psi)\=1  \~.\label{jacid02b}
\eeq
{\sc Proof of \eq{jacid02b}}: 
\bea
&&6 \llbracket\llbracket\Psi,\Psi\rrbracket,\Psi\rrbracket
- D[\llbracket\Psi,\Psi\rrbracket, \Psi]
- \Hi \epsilon\q_{ab} [\Delta^{a},[[\Delta^{b},\Psi],
\llbracket\Psi,\Psi\rrbracket]] \cr
&=& 3[\llbracket\Psi,\Psi\rrbracket,D\Psi] 
+ 3[D\llbracket\Psi,\Psi\rrbracket,\Psi] 
- D[\llbracket\Psi,\Psi\rrbracket,\Psi] 
- \Hi \epsilon\q_{ab} [\Delta^{a},[[\Delta^{b},\Psi],
\llbracket\Psi,\Psi\rrbracket]] \cr
&\equi{\e{weird01}}& 
4[[D\Psi,\Psi],D\Psi] + 2 [D[D\Psi,\Psi],\Psi]  
\= -4[D\Psi,[D\Psi,\Psi]] + 2 [[D\Psi,D\Psi],\Psi]\=0 \~. \label{jacid02c}
\eea

\section{Space of Solutions to QME}
\label{secsolqme}

\noi
We can generate a new solution to the QME \e{qmew} via a finite transformation
\beq
w\~\longrightarrow\~w^{\prime}\= (e^{D\Psi} w)\~, \qquad 
\eps(\Psi)\=1\~, \qquad \pl(\Psi)\~\geq \~0\~,\label{newsol01}
\eeq
where $D$ is the Grassmann-odd $1$-quantum antibracket \e{dee01}.
The composition of two finite transformations is again a finite transformation
\beq
e^{D\Psi\q_{1}} e^{D\Psi\q_{2}}
\= e^{{\rm BCH}(D\Psi\q_{1},D\Psi\q_{2})}
\= e^{D{\rm BCH}(\Psi\q_{1},\Psi\q_{2})}\~,\qquad \eps(\Psi\q_{i})\=1\~. \label{bch01a}
\eeq
The second and third expression in \eq{bch01a} contain the 
Baker--Campbell--Hausdorff (BCH) series expansion (with the Lie bracket 
replaced with the commutator $[\cdot,\cdot]$ and the $2$-quantum antibrackets 
$\llbracket\cdot,\cdot\rrbracket$, respectively). In detail, the latter reads
\bea
{\rm BCH}(\Psi\q_{1},\Psi\q_{2})
&=&\Psi\q_{1}+\int_{0}^{1}\! dt \sum_{n=0}^{\infty}\frac{(-1)^{n}}{n+1}
\left(e^{-t\llbracket\Psi\q_{2},\~\cdot\~\rrbracket}\~
e^{-\llbracket\Psi\q_{1},\~\cdot\~\rrbracket}-1\right)^{n}\Psi_{2} \cr
&=&\Psi\q_{1}+\Psi\q_{2}
+\Hf\llbracket\Psi\q_{1},\Psi\q_{2}\rrbracket+
\frac{1}{12}\llbracket\Psi\q_{1},\llbracket\Psi\q_{1},
\Psi\q_{2}\rrbracket\rrbracket
+\frac{1}{12}\llbracket\llbracket\Psi\q_{1},\Psi\q_{2}\rrbracket,
\Psi\q_{2}\rrbracket
+{\cal O}(\Psi_{i}^{4})\~. \label{bch01b}
\eea
Here we have used the Jacobi identity \e{jacid01a}.

\section{$Sp(2)$ case}
\label{secsolqme2}

\noi
There is an $Sp(2)$-symmetric analogue of Section~\ref{secsolqme}.
We can generate a new solution via the finite transformation
\beq
w\~\longrightarrow\~w^{\prime}\= (e^{DB} w)\~, \qquad 
\eps(B)\=0\~, \qquad \pl(B)\~\geq \~0\~. \label{newsol02}
\eeq
where $D$ is the Grassmann-even $1$-quantum bracket \e{dee02}. 
The composition of two finite transformations is again a finite transformation
\beq
e^{DB\q_{1}} e^{DB\q_{2}}
\= e^{D{\rm BCH}(B\q_{1},B\q_{2})}\~,\qquad \eps(B\q_{i})\=0\~, \label{bch02a}
\eeq
with a BCH formula in \eq{bch02a} for the bosons $B\q_{i}$ similar to the 
formula \e{bch01b} for the fermions $\Psi\q_{i}$.

\section{Maximal Deformation}
\label{secmaxded}

\noi
One may formally argue \cite{bt94def} that any two solutions to the QME 
\e{qmew} are connected via a finite transformation \e{newsol01}, \ie the 
group of finite transformations \e{newsol01} acts transitively on the space of 
solutions to the QME \e{qmew}.

\noi
The infinitesimal generator of an infinitesimal transformation \e{newsol01} 
\beq
\delta w\=\left( [\Delta,\Psi]w \right)\~
\equi{\e{qmew}}\~(\Delta\Psi w) \~,\label{deltaw1a}
\eeq
is evidently the $1$-antibracket $D\Psi\equiv[\Delta,\Psi]$ for an 
infinitesimal operator $\Psi$ with $\eps(\Psi)=1$ and $\pl(\Psi)\geq 0$.
Phrased equivalently, \eq{deltaw1a} means that 
the change in the master action is given by the quantum BRST transformation 
\beq
  \delta W\= \Hi \delta\ln w \= \sigma\q_{W}\Psi\~.\label{deltaw1b}
\eeq
The same story holds for $X$ instead of $W$ if we replace the operator
$\Delta$ with the transposed operator $\Delta^{T}$, \eg
\beq
\delta x\=\left( [\Delta^{T},\Psi]x\right)
\~\equi{\e{qmex}}\~(\Delta^{T}\Psi x) \~, \qquad  
\delta X\= \Hi \delta\ln x \= \sigma\q_{X}\Psi\~.\label{deltax1}
\eeq
When discussing $X$ (as opposed to $W$) we will implicitly assume that the 
pertinent quantum (anti)brackets from Section~\ref{sechighqantibrackets}
are generated by the transposed operator 
$\Delta^{T}$. 

\noi
Moreover, to obtain the $Sp(2)$-symmetric formulation, formally replace the 
operator $\Delta\to\Delta^{a}$ and 
$\Psi\to \Psi\q_{a}\equiv\Hf\Hi\epsilon\q_{ab}[\Delta^{b},B]$. Note that 
$\pl(\Psi\q_{a})\geq 0$ holds.

\section{Gauge-Independence via Integration by Parts}
\label{secgaugeindepintbyparts}

\noi
The gauge-independence of the path integral can be formally proved via
integration by parts
\beq
\delta Z \~\equiv\~ Z\p_{X+\delta X} - Z\p_{X} 
\~\equi{\e{pathint01}}\~ \int \! d\mu \~w \~\delta x
\~\equi{\e{deltax1}}\~ \int \! d\mu \~w (\Delta^{T}\Psi x) 
\~\equi{{\rm int.\ by~parts}}\~ \int \! d\mu \~(\Delta w)\~ (\Psi x) 
\~\equi{\e{qmew}}\~0\~.\label{deltapathint01} 
\eeq
Eq.\ \e{deltapathint01} is the main result of \Ref{bbd96}.
The main purpose of this paper is to re-prove gauge-independence via change of 
variables in the path integral, \cf Section~\ref{secgaugeindepchange}. To 
this end, we introduce two types of homotopy operators, \cf 
Sections~\ref{secho}--Sections~\ref{secbho}.

\section{Homotopy Operator $\larrow{h}^{A}\!\!(\Delta)$}
\label{secho}

\noi
The pertinent {\bf homotopy operator} $\larrow{h}^{A}\!\!(\Delta)$ is best 
explained for operators $\Delta$ on antinormal-ordered form
\beq
\Delta(\partial,z) 
\=\sum_{m=0}^{\infty}\Delta\q_{m}(\partial,z) \~, \qquad  
\Delta\q_{m}(\partial,z)\= 
\larrow{\partial}\pp_{A\q_{m}}\ldots 
\larrow{\partial}\pp_{A\q_{1}} 
\Delta_{m}^{A\q_{1}\ldots A\q_{m}}(z)\~. \label{deltaantinormal}
\eeq
We stress that the derivatives $\larrow{\partial}\pp_{A\q_{m}}\ldots 
\larrow{\partial}\pp_{A\q_{1}}$ in \eq{deltaantinormal} also act beyond 
(\ie to the right of) $\Delta_{m}^{A\q_{1}\ldots A\q_{m}}(z)$. Then the homotopy 
operator is defined on a homogeneous component $\Delta\q_{m}(\partial,z)$ as
\beq
\larrow{h}^{A}\!\!(\Delta\q_{m})
\~:=\~ \left\{ \begin{array}{lcl} 
-\frac{1}{m}[z^{A}, \Delta_{m}] \= (-1)^{\eps_{A}} 
\larrow{\partial}\pp_{A\q_{m-1}}\ldots 
\larrow{\partial}\pp_{A\q_{1}} 
\Delta_{m}^{A\q_{1}\ldots A\q_{m-1}A}(z)
&\for & m\geq 1\~, \\ \\ 0&\for & m=0\~. \end{array} \right.\label{ho01}
\eeq
The definition \e{ho01} is extended to an arbitrary operator $\Delta$ by 
linearity. The homotopy operator \e{ho01} satisfies the following homotopy 
property
\beq
(-1)^{\eps_{A}} \larrow{\partial}\pp_{A}\larrow{h}^{A}\!\!(\Delta(\partial,z)) 
\= \Delta(\partial,z) - \Delta(0,z)  \label{ho02}
\eeq
for antinormal-ordered operators \e{deltaantinormal}.
Two homotopy operators \e{ho01} commute:
\beq
\larrow{h}^{A} \larrow{h}^{B}\!\!(\Delta) 
\= (-1)^{\eps_{A}\eps_{B}} \larrow{h}^{B} \larrow{h}^{A}\!\!(\Delta) \~. \label{ho03}
\eeq

\section{Bilinear Homotopy Operator $B^{A}(f,\Delta)$}
\label{secbho}

\noi
Given a function $f$ and an operator $\Delta$, the {\bf bilinear homotopy 
operator} $B^{A}(f,\Delta)$ is defined via
\bea
(-1)^{\eps_{A}\eps_{f}}B^{A}(f,\Delta) &\equiv& 
f\~:\frac{1}{1 -\rarrow{\partial}^{T}_{B}\~\larrow{h}^{B}}:
\~\larrow{h}^{A}\!\!(\Delta)1
\~\equiv\~ 
f\~:\sum_{n=0}^{\infty}\left(\rarrow{\partial}^{T}_{B}
\~\larrow{h}^{B}\right)^{n}:\~\larrow{h}^{A}\!\!(\Delta)1 \cr
&\equiv& 
f \~\larrow{h}^{A}\!\!(\Delta)1 
+(f\rarrow{\partial}^{T}_{B})\~\larrow{h}^{B}\larrow{h}^{A}\!\!(\Delta)1 
+(f\rarrow{\partial}^{T}_{B}\rarrow{\partial}^{T}_{C})
\~\larrow{h}^{C}\larrow{h}^{B}\larrow{h}^{A}\!\!(\Delta)1 +\ldots\~, 
\label{bho01}
\eea
where the ordering symbol ``$:\~:$'' here means that all derivatives 
$\rarrow{\partial}^{T}_{B}$ should be to the left of all the homotopy operators
$\larrow{h}^{B}$. One may prove that the bilinear homotopy operator 
$B^{A}(f,\Delta)$ has the following important homotopy property
\beq
(-1)^{\eps_{A}} \left( \partial^{T}_{A}B^{A}(f,\Delta) \right)
\= (-1)^{\eps_{f}\eps_{\Delta}} (\Delta^{T} f)-f (\Delta 1) \~. \label{bho02} 
\eeq

\section{Gauge-Independence via Change of Variables}
\label{secgaugeindepchange}

\noi
The infinitesimal change $\delta z^{A}$ of (passive) coordinates $z^{A}$ can 
be viewed as an infinitesimal vector field\footnote{We are here and below 
guilty of infusing some active picture language into a passive picture, 
\ie properly speaking, the active vector field has the opposite sign.}
\beq
\delta z^{A} \= \frac{1}{wx} B^{A}( \Psi x, \larrow{\Delta} w)\~, \qquad 
\pl(\Psi)\~\geq\~0\~. 
\label{deltaz01}
\eeq
One may show that the Planck number $\pl( \delta z^{A})\geq -1 $ of the 
vector field is greater than or equal to $-1$, as it should be. 
The Boltzmann density (= integrand) of the path integral \e{pathint01} 
is $\rho wx$. The divergence of the vector field \e{deltaz01} \wrt the 
Boltzmann density is
\bea
\div\q_{\rho  wx} \delta z &\equiv&
\frac{(-1)^{\eps_{A}}}{\rho  wx}(\larrow{\partial}\pp_{A} \rho x w \~\delta z^{A}) 
\~\equi{\e{deltaz01}}\~ 
-\frac{(-1)^{\eps_{A}}}{wx}(\larrow{\partial}^{T}_{A}
B^{A}( \Psi x, \larrow{\Delta} w) ) \cr
&\equi{\e{bho02} }& \frac{1}{wx}
\left\{(\Psi x) (\Delta w) + w (\Delta^{T}\Psi x) \right\} \cr
&\equi{\e{qmew}+\e{qmex}}& \frac{1}{x} ([\Delta^{T},\Psi]x) 
\~\equi{\e{sigmax01}}\~ \Ih\sigma\q_{X}\Psi
\~\equi{\e{deltax1}}\~\frac{\delta  x}{x} \~.\label{divrhoxw01}
\eea
On one hand, an infinitesimal change of integration variables in path 
integral cannot change the value of path integral. On the other hand, 
it induces an infinitesimal Jacobian factor. Hence 
\beq
 0\= \int \! [d\lambda][dz]\~
(-1)^{\eps_{A}}(\larrow{\partial}\pp_{A} \rho x w\~\delta z^{A}) 
\= \int \! d\mu\~ wx \~\div\q_{\rho x w} \delta z
\~\equi{\e{divrhoxw01}}\~ \int \! d\mu\~ w\~\delta x
\~\equi{\e{pathint01}}\~Z\p_{X+\delta X} - Z\p_{X}\~\equiv\~\delta Z  \~,
\label{deltapathint02} 
\eeq
which, in turn, can mimic an arbitrary infinitesimal change of gauge-fixing. 
Thus we have formally proven via change of variables that the path integral 
$Z\p_{X}$ does not depend on gauge-fixing $X$. Eq.\ \e{deltapathint02} is the 
main result of this article.

\section{Higher antibrackets}
\label{sechighcantibrackets}

\noi 
The {\bf $n$-antibracket} \cite{bda96,bbd96,b06cmp} is the restriction of 
the quantum $n$-antibracket \e{phin01} from operators to functions 
\beq
\Phi_{\Delta}^{n}(f\q_{1},\ldots, f\q_{n}) \~:=\~
\Hat{\Phi}_{\Delta}^{n}(f\q_{1},\ldots, f\q_{n})1\~. \label{phin01fct}
\eeq
In particular, the {\bf $2$-antibracket} $(f,g)$ of two functions $f$ and $g$
is defined as 
\beq
(f,g)\~:=\~(-1)^{\eps_{f}} \left[\left[\larrow{\Delta},f \right],g \right]1
\=-\llbracket f,g \rrbracket 1   \= -(-1)^{(\eps_{f}+1)(\eps_{g}+1)}(g,f)\~.
\label{antibracket01}
\eeq

\section{Second-Order $\Delta$ operator}
\label{sec2ndorder}

\noi
It is natural to ponder how to build a nilpotent $\Delta$-operator, that
takes scalar functions in scalar functions, from the following given
geometric data:

\begin{enumerate}
\item
An {\bf anti-Poisson structure}
\beq
(f,g)\= (f\rarrow{\partial}\pp_{A}) E^{AB} (\larrow{\partial}\pp_{B}g)
\=-(-1)^{(\eps_{f}+1)(\eps_{g}+1)}(g,f)\~, \quad 
\eps( E^{AB} )\= \eps_{A}+ \eps_{B}+1\~, \quad \pl(E^{AB})\geq 0\~, 
\label{antibracket06}
\eeq
which satisfies the Jacobi identity 
\beq
\sum_{f,g,h~{\rm cycl.}}(-1)^{(\eps_{f}+1)(\eps_{h}+1)}(f,(g,h))\=0\~. \label{jacid00}
\eeq

\item
A {\bf density} $\rho$ with $\eps(\rho)=0$ and $\pl(\ln\rho)\geq -1$.
\item
A Grassmann-odd {\bf vector field} $V=V^{A}\larrow{\partial}\pp_{A}$, with 
$\eps(V)=1$ and $\pl(V)\geq -2$, that is compatible with the
anti-Poisson structure:
\beq
(V(f,g))  \= (Vf,g) - (-1)^{\eps_{f}}(f,Vg) \~.\label{veecompatiple06}
\eeq
\end{enumerate}

\noi
Often we assume that the antibracket \e{antibracket06} is 
non-degenerate/invertible.
Then the vector field is locally a Hamiltonian vector field $V=(H,\cdot)$.
This Hamiltonian $H$ can be absorbed into the density by redefining the density 
$\Tilde{\rho}=\rho e^{2H}$.

\noi
To guarantee nilpotency $\Delta^{2}=0$, the minimal solution (to the above 
posed problem in Section~\ref{sec2ndorder}) is the following second-order
$\Delta$ operator
\beq
\Delta\= \Delta\q_{\rho} + V + \nu \~, 
\qquad \eps(\Delta )\= 1\~, \qquad \pl(\Delta)\~\geq\~ -2\~,
\label{delta2ndorder01}
\eeq
where $\Delta\q_{\rho}$ is the odd Laplacian
\beq
\Delta\q_{\rho} \= \frac{(-1)^{\eps_{A}}}{2\rho}\larrow{\partial}\pp_{A} \rho E^{AB}
\larrow{\partial}\pp_{B}
\=-\frac{(-1)^{\eps_{A}}}{2}\larrow{\partial}^{T}_{A} E^{AB}
\larrow{\partial}\pp_{B}\~,\label{oddlapl01}
\eeq
where $\nu$ is an odd scalar function 
\beq
\nu \= \nu\p_{\rho} +\hf\div\p_{\rho}V - \hf V^{A}E\q_{AB}V^{B}\~, 
\qquad \eps(\nu)\= 1\~, \qquad \pl(\nu)\~\geq\~-2\~,\label{nu01}
\eeq
and where the odd scalar $\nu\p_{\rho}$ is constructed from $\rho$ and $E^{AB}$,
\cf Refs.\ \cite{b06,b07,bb07,bb08,bb09}. The transposed vector field is
\beq
V^{T} \=-V-\div\p_{\rho}V\~.\label{veetee01}
\eeq
The transposed operator $\Delta^{T}$ corresponds to letting the vector field 
$V\to-V$ change sign: 
\beq
\Delta^{T}\= \left. \Delta \right|_{V\to -V} \~.\label{delta2ndordertee01}
\eeq

\noi
To obtain the $Sp(2)$-symmetric formulation, formally replace 
$\Delta\q_{\rho} \to\Delta^{a}_{\rho}$; $(\cdot,\cdot) \to(\cdot,\cdot)^{a}$; 
$V\to V^{a}$; $\nu\to \nu^{a}$; etc. Note that some equations, such as, \eg 
\es{jacid00}{veecompatiple06} should be symmetrized in the $Sp(2)$ indices.
We will not here discuss an $Sp(2)$-analogue of \eq{nu01}.

\section{Application to the Second-Order $\Delta$ operator}
\label{sec2ndorderappl}

\noi
Now let us check how the higher-order formalism of the previous 
Sections~\ref{secdeltaop}-\ref{secgaugeindepchange} applies to the 
second-order $\Delta$ operator \e{delta2ndorder01}.
The QME \e{qmew} becomes 
\beq
\Hf (W,W) +  \hi\left(\left( \Delta\q_{\rho} +V\right)W\right) 
+ \left(\hi\right)^2 \nu \= 0\~,\label{qmew02}
\eeq
and the BRST operator \e{sigmaw01} becomes
\beq
 \sigma\q_{W} f \=\hi\left(\left(\Delta\q_{\rho} +V\right)f\right) +(W,f)\~. 
\label{sigmaw02}
\eeq
The homotopy operator \e{ho01} becomes
\beq
\larrow{h}^{A}\!\!(\larrow{\Delta}w)\= \Hf E^{AB} \larrow{\partial}\pp_{B}w 
+ (-1)^{\eps_{A}} V^{A} w +(z^{A},\ln\sqrt{\rho})w\~,\label{ho2nd01}
\eeq
\beq
\larrow{h}^{B}\larrow{h}^{A}\!\!(\larrow{\Delta}w)
\= \frac{(-1)^{\eps_{A}}}{2} E^{AB}w\~.\label{ho2nd02}
\eeq
The infinitesimal change \e{deltaz01} of variables becomes
\beq
2x\~\delta z^{A}\~\equi{\e{deltaz01}}\~
\frac{2}{w} B^{A}( \Psi x, \larrow{\Delta} w)
\= (\Psi x) \left( (\ln w, z^{A})+ 2V^{A}\right) - (\Psi x,z^{A})\~,
\label{deltaz2nd01}
\eeq
where $\Psi$ is an infinitesimal operator. For an infinitesimal function 
$\psi$, \eq{deltaz2nd01} reduces further to
\beq
2\~\delta z^{A}\equi{\e{deltaz2nd01}}\~
\Ih \psi \~(W-X, z^{A}) +2\psi V^{A}  - (\psi,z^{A})
\= \Ih \psi\~( \sigma\q_{W}z^{A} - \sigma\q_{X}z^{A})  
- (\psi,z^{A})\~.
\label{deltaz2nd02}
\eeq
{}Finally, consider a finite change of solution to the QME \e{qmew}
\beq
 w^{\prime} \~\equiv\~ e^{\Ih W^{\prime}} 
\= \left(e^{-D\psi} w\right) \~,\qquad 
D\psi\~\equi{\e{dee01}}\~[\Delta,\psi]\=(\Delta\psi)- \ad\psi\~,\qquad 
\ad\psi \~\equiv\~(\psi,\cdot)\~,
\eeq 
where $\psi$ is a finite function, with $\eps(\psi)=1$ and $\pl(\psi)\geq 0$. 
An application of the BCH formula shows that the corresponding change 
in the action reads \cite{bbd06,blt14lagr1,bblt14lagr2}
\beq
 W^{\prime}\=e^{\ad\psi} W + i\hbar\~E(\ad\psi)(\Delta \psi)
\=e^{\ad\psi} W + i\hbar\frac{e^{\ad\psi}-1}{\ad\psi}(\Delta \psi)
\~, \qquad   w^{\prime}\= (e^{\ad\psi}w)\~e^{-E(\ad\psi)(\Delta \psi)}\~,
\label{newsol2nd}
\eeq
where
\beq
E(x)\~:=\~\int_{0}^{1}\! dt\~ e^{xt}\= \frac{e^{x}-1}{x}\~.\label{ernoulli01}
\eeq
{\sc Proof of \eq{newsol2nd}}:
For a vector field $\xi$ and a function $f$, the BCH formula simplifies to
\beq
e^{\xi}e^{f} \= e^{\xi+ B(-[\xi,\cdot])f}\~,\label{bernoulli01}
\eeq
where 
\beq
B(x)\~:=\~ \frac{x}{e^{x}-1}\=\frac{1}{E(x)}
\= 1-\frac{x}{2}+\frac{x^{2}}{12}-\frac{x^{4}}{720}+{\cal O}(x^{6})
\label{bernoulli02}
\eeq
is the generating function for Bernoulli numbers. Therefore \eq{bernoulli01}
can be inverted into
\beq
e^{\xi+f} \= e^{\xi}e^{E(-[\xi,\cdot])f}\~,\label{ernoulli02}
\eeq
which, in turn, leads to \eq{newsol2nd} with $\xi=\ad\psi$ and 
$f=-(\Delta\psi)$.

\vspace{0.8cm}

\noi
{\sc Acknowledgement:}~The authors would like to thank Poul H.\ Damgaard, 
Peter M.\ Lavrov and Igor V.\ Tyutin for interesting discussions. 
K.B.\ would like to thank K.P.~Zybin and the Lebedev Physics Institute for
warm hospitality. The work of I.A.B.\ is supported in part by the RFBR grants
14-01-00489 and 14-02-01171. The work of K.B.\ is supported by the Grant
Agency of the Czech Republic (GACR) under the grant P201/12/G028.


\begin{thebibliography}{999}

\bibitem{bbd96} 
I.A.~Batalin, K.~Bering and P.H.~Damgaard, Phys.~Lett.\ {\bf B389} (1996) 673, 
arXiv:hep-th/9609037.


\bibitem{witten90} E.~Witten, Mod.~Phys.~Lett.\ {\bf A5} (1990) 487.


\bibitem{bv81} 
I.A.~Batalin and G.A.~Vilkovisky, Phys.~Lett.\ {\bf 102B} (1981) 27.

\bibitem{bv83} I.A.~Batalin and G.A.~Vilkovisky,
Phys.~Rev.\ {\bf D28} (1983) 2567 [E: {\bf D30} (1984) 508].

\bibitem{bv84} 
I.A.~Batalin and G.A.~Vilkovisky, Nucl.~Phys.\ {\bf B234} (1984) 106.


\bibitem{bltla2} I.A.~Batalin, P.~Lavrov and I.~Tyutin, 
J.~Math.~Phys.\ {\bf 31} (1990) 1487; 
{\em ibid.} {\bf 32} (1991) 532; 
{\em ibid.} {\bf 32} (1991) 2513.


\bibitem{bt94def}
I.A.~Batalin and I.V.~Tyutin, Int.~J.~Mod.~Phys.\ {\bf A9} (1994) 517.


\bibitem{bt92} I.A.~Batalin and I.V.~Tyutin,
Int.~J.~Mod.~Phys.\ {\bf A8} (1993) 2333, arXiv:hep-th/9211096.

\bibitem{bt93} I.A.~Batalin and I.V.~Tyutin, 
Mod.~Phys.~Lett.\ {\bf A8} (1993) 3673, arXiv:hep-th/9309011.

\bibitem{bt94} I.A.~Batalin and I.V.~Tyutin, 
Mod.~Phys.~Lett.\ {\bf A9} (1994) 1707, arXiv:hep-th/9403180. 
 
\bibitem{bms95} I.A.~Batalin, R.~Marnelius and A.M.~Semikhatov, 
Nucl.~Phys.\ {\bf B446} (1995) 249, arXiv:hep-th/9502031.

\bibitem{bm96} I.A.~Batalin and R.~Marnelius, 
Nucl.~Phys.\ {\bf B465} (1996) 521, arXiv:hep-th/9510201.

\bibitem{bt96} I.A.~Batalin and I.V.~Tyutin,
Amer.~Math.~Soc.~Transl.\ {\bf 2.177} (1996) 23.

\bibitem{bbd06} I.A.~Batalin, K.~Bering and P.H.~Damgaard, 
Nucl.~Phys.\ {\bf B739} (2006) 389, arXiv:hep-th/0512131.


\bibitem{bv85} 
I.A.~Batalin and G.A.~Vilkovisky, J.~Math.~Phys.\ {\bf 26} (1985) 172.

\bibitem{bb10} I.A.~Batalin and K.~Bering, 
Int.~J.~Mod.~Phys.\ {\bf A25} (2010) 2119, arXiv:0911.0341.


\bibitem{k99} O.M.~Khudaverdian, arXiv:math.DG/9909117.

\bibitem{kv02} O.M.~Khudaverdian and Th.~Voronov, 
Lett.~Math.~Phys.\ {\bf 62} (2002) 127, arXiv:math.DG/0205202. 

\bibitem{k02} O.M.~Khudaverdian, 
Contemp.~Math.\ {\bf 315} (2002) 199, arXiv:math.DG/0212354.

\bibitem{k04} O.M.~Khudaverdian, 
Commun.~Math.~Phys.\ {\bf 247} (2004) 353, arXiv:math.DG/0012256.

\bibitem{b06} 
K.~Bering, J.~Math.~Phys.\ {\bf 47} (2006) 123513, arXiv:hep-th/0604117.

\bibitem{b07}
K.~Bering, J.~Math.~Phys.\ {\bf 49} (2008) 043516, arXiv:0705.3440.

\bibitem{bb07} I.A.~Batalin and K.~Bering, 
J.~Math.~Phys.\ {\bf 49} (2008) 033515, arXiv:0708.0400.

\bibitem{bb08} I.A.~Batalin and K.~Bering,
Phys.~Lett.\ {\bf B663} (2008) 132, arXiv:0712.3699.

\bibitem{bb09} I.A.~Batalin and K.~Bering, J.~Math.~Phys.\ {\bf 50} 
(2009) 073504, arXiv:0809.4269.


\bibitem{yks96}
Y.~Kosmann-Schwarzbach, Ann.~Inst.~Fourier (Grenoble) {\bf 46} (1996) 1243.

\bibitem{yks04} Y.~Kosmann-Schwarzbach, Lett.~Math.~Phys.\ {\bf 69} (2004) 61, 
arXiv:math/0312524.

\bibitem{bm98}
I.A.~Batalin and R.~Marnelius, Phys.~Lett.\ {\bf B 434} (1998) 312, 
arXiv:hep-th/9805084.

\bibitem{bm99}
I.A.~Batalin and R.~Marnelius, Nucl.~Phys.\ {\bf B551} (1999) 450,
arXiv:hep-th/9809208.

\bibitem{bm99dual} 
I.A.~Batalin and R.~Marnelius, Int.~J.~Mod.~Phys.\ {\bf A14} (1999) 5049, 
arXiv:hep-th/9809210.

\bibitem{bm99nonilp} 
I.A.~Batalin and R.~Marnelius, Theor.~Math.~Phys.\ {\bf 120} (1999) 1115, 
arXiv:hep-th/9905083.

\bibitem{b06cmp} 
K.~Bering, Commun.~Math.~Phys.\ {\bf 274} (2007) 297, 
arXiv:hep-th/0603116.


\bibitem{bda96}
K.~Bering, P.H.~Damgaard and J.~Alfaro, Nucl.~Phys.\ {\bf B478} (1996) 459,
arXiv:hep-th/9604027.

\bibitem{bbd97} 
I.A.~Batalin, K.~Bering and P.H.~Damgaard, Phys.~Lett.\ {\bf B408} (1997) 235, 
arXiv:hep-th/9703199.


\bibitem{blt14lagr1} 
I.A.~Batalin, P.M.~Lavrov and I.V.~Tyutin, 
Int.~J.~Mod.~Phys.\ {\bf A29} (2014) 1450166, arXiv:1405.2621.

\bibitem{bblt14lagr2} 
I.A.~Batalin, K.~Bering, P.M.~Lavrov and I.V.~Tyutin, 
Int.~J.~Mod.~Phys.\ {\bf A29} (2014) 1450167, arXiv:1406.4695.

\end{thebibliography}
\end{document}